\documentclass[conference]{IEEEtran}
\IEEEoverridecommandlockouts

\usepackage{cite}
\usepackage{amsmath,amssymb,amsfonts}
\usepackage{algorithmic}
\usepackage{graphicx}
\usepackage{textcomp}
\usepackage{xcolor}
\def\BibTeX{{\rm B\kern-.05em{\sc i\kern-.025em b}\kern-.08em
    T\kern-.1667em\lower.7ex\hbox{E}\kern-.125emX}}

\usepackage{bm} 
\usepackage{esint} 
\usepackage{commath} 
\usepackage{nicefrac} 
\usepackage{subcaption}  

\usepackage{booktabs}

\usepackage{enumitem} 
\setlist{nolistsep} 

\usepackage{multirow} 
\usepackage{rotating} 

\usepackage[unicode=true,pdfusetitle,%
 pdfauthor={Shivam Mehta, Nebojsa Jojic, Hannes Gamper},%
 pdfkeywords={audio language models, multimodal LLMs, audio reasoning, audio captioning, audio tokenization, audio generation},%
 bookmarks=true,bookmarksnumbered=true,bookmarksopen=false,%
 breaklinks=true,pdfborder={0 0 0},backref=false,colorlinks=true,citecolor=blue]{hyperref}

\newcommand{\architectureNameWithAbbrevation}{Language Model-Make Some Noise (LM-MSN)}
\newcommand{\architectureAbbrevationOnly}{LM-MSN}

\newcommand{\real}{\mathbb{R}}

\newcommand{\x}{\boldsymbol{x}}

\begin{document}
\title{Make Some Noise: Towards LLM audio reasoning and generation using sound tokens\\
\thanks{\textsuperscript{*}All work performed during internship at Microsoft Research.}
}

\author{\IEEEauthorblockN{Shivam Mehta\textsuperscript{*}}
\IEEEauthorblockA{
    \textit{Division of Speech, Music and Hearing} \\ \textit{KTH Royal Institute of Technology, Sweden} \\ {smehta@kth.se}
}
\and    
\IEEEauthorblockN{Nebojsa Jojic}
\IEEEauthorblockA{
    \textit{Microsoft Research Redmond} \\ {Washington, USA} \\
    {jojic@microsoft.com}
}
\and
\IEEEauthorblockN{Hannes Gamper}
\IEEEauthorblockA{
    \textit{Microsoft Research Redmond} \\ {Washington, USA} \\
    {hannes.gamper@microsoft.com}    
}
}

\maketitle

\begin{abstract}
Integrating audio comprehension and generation into large language models (LLMs) remains challenging due to the continuous nature of audio and the resulting high sampling rates. Here, we introduce a novel approach that combines Variational Quantization with Conditional Flow Matching to convert audio into ultra-low bitrate discrete tokens of 0.23kpbs, allowing for seamless integration with text tokens in LLMs. We fine-tuned a pretrained text-based LLM using Low-Rank Adaptation (LoRA) to assess its effectiveness in achieving true multimodal capabilities, i.e., audio comprehension and generation. Our tokenizer outperforms a traditional VQ-VAE across various datasets with diverse acoustic events. Despite the substantial loss of fine-grained details through audio tokenization, our multimodal LLM trained with discrete tokens achieves competitive results in audio comprehension with state-of-the-art methods, though audio generation is poor. Our results highlight the need for larger, more diverse datasets and improved evaluation metrics to advance multimodal LLM performance.
\end{abstract}


\section{Introduction}
In recent years, significant advancements have been made in the development of multimodal large language models (MLLM) capable of multimodal understanding and generation~\cite{liu2024visual, zhang2023video, ge2023making, team2024chameleon, yu2023scaling}. However, creating a unified framework for audio comprehension and generation continues to be a significant challenge. Current deep neural network approaches to audio processing have achieved remarkable success in various domains, including audio representation learning~\cite{elizalde2022clap,wu2022large, baevski2020wav2vec, hsu2021hubert}, classification~\cite{huang2022masked, gong2022ssast, srivastava2022conformer, chen2022hts_at}, generation~\cite{audioldm2, agostinelli2023musiclm, musicgen, kreuk2022audiogen} and most recently, audio understanding~\cite{kong2024audio, gong2023ltuas, gong2023listen, deshmukh2023pengi}. Traditional classification models~\cite{huang2022masked, gong2022ssast, srivastava2022conformer, chen2022hts_at} have primarily focused on learning continuous audio representations and mapping them to predefined discrete sound labels, achieving considerable success in tasks such as audio event recognition and classification. Furthermore, to enhance open-domain comprehension and reasoning, these continuous audio representations are provided to Large Language Models (LLMs) \cite{wu2022large, touvron2023llama} along with text inputs. Recent works \cite{kong2024audio, gong2023ltuas, gong2023listen} have highlighted the potential of this approach and offer a promising avenue for enhancing audio understanding beyond classification.

However, most existing large audio language models rely on a continuous representation of audio \cite{baevski2020wav2vec, gong2023ltuas, chu2023qwen, kong2024audio}. While these continuous representations offer fine-grained information that is advantageous for tasks involving audio comprehension, they present significant limitations in the context of audio generation. The core challenge lies in the training paradigm of LLMs, which are typically trained using a next-token prediction task. LLMs excel in this task because they operate within a discrete token space, where each token represents a distinct, quantized unit of information. In contrast, continuous audio representations occupy a smooth, high-dimensional latent space that does not naturally align with the discrete token prediction framework required for LLM training. As a result, the continuous nature of these representations hinders the effective training of audio LLMs for generation tasks.

\begin{figure*}[t]
    \centering
    \includegraphics[width=\textwidth]{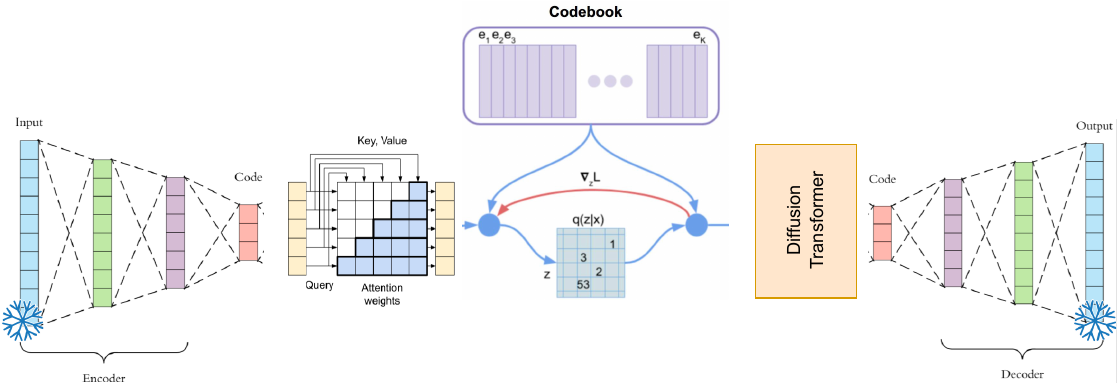}
    \caption{Architecture of audio tokenizer containing frozen autoencoder follow by a causal encoder and a conditional flow matching-based decoder with Diffusion Transformer to reconstruct representations from quantised vectors.}
    \label{fig: quantizer architecture}
    \vspace{-10pt}
\end{figure*}

To achieve unified audio comprehension and generation, we encode continuous audio to discrete tokens for LLM training. By quantizing audio into tokens with low time resolution, we align audio and text token counts, making transformer training feasible given its quadratic complexity. This minimizes vocabulary size and temporal resolution, enabling efficient multimodal training.
While this process involves some sacrifice of fine-grained detail, it is a necessary step toward developing a unified model that can seamlessly handle both audio comprehension and generation tasks. 

Building on this approach, this paper introduces \architectureNameWithAbbrevation, a novel methodology designed to fuse audio tokens with text tokens, thereby paving the way for unified audio reasoning and generation within LLMs using discrete sound tokens. Our work makes two primary contributions:
\begin{itemize}
    \item We propose a novel Variational Quantization (VQ) \cite{van2017neural} mechanism that leverages diffusion transformers \cite{peebles2023scalable} and Conditional Flow Matching (CFM) \cite{lipman2023flow} to effectively tokenize audio waveforms to ultra-low bitrate discrete audio tokens (0.23 kbps).

    \item We propose a pipeline to fine-tune LLMs for unified audio comprehension and generation via early fusion of these quantized audio representations with text tokens using LoRA \cite{hu2021lora}.
\end{itemize}

This early methodology marks a significant step toward integrating audio processing into the broader framework of multimodal LLMs.

\section{Background}
\subsection{Audio Tokenization}
Audio waveforms are typically sampled at high rates of up to 48 kHz, with the sampling rate representing the number of data points captured per second of audio. The high number of data points generated at these sampling rates can be computationally intensive for various applications, often necessitating downsampling or compression. One effective method of reducing the data rate is through the quantization of audio signals \cite{hsu2021hubert, chiu2022self, van2017neural}, a process often achieved using a discrete bottleneck. Recent approaches \cite{zeghidour2021soundstream, defossez2022high_encodec, kumar2024high_descript} have employed Residual Vector Quantization (RVQ) and adversarial losses to generate low-bitrate vector codebooks, while others have used VQ-VAE alongside powerful decoders such as WaveNet \cite{van2016wavenet} to reconstruct audio from these compressed representations \cite{garbacea2019low}. However, there has been limited exploration of modifying the reconstruction objectives using other probabilistic models including diffusion or flow matching.



\subsection{Conditional Flow Matching}

Conditional Flow Matching (CFM or FM)  \cite{lipman2023flow, mehta2024matcha} presents a simulation-free approach for learning a vector field $\boldsymbol{u}_t: [0, 1] \times \real^d \rightarrow \real^d$, where $t \in [0, 1]$ that maps probability path $p_t$ between samples from $p_0(\x) \sim \mathcal{N}(\boldsymbol{0}, \boldsymbol{I})$ and samples $p_1(\x)$, where $p_1(\x)$ approximates the unknown data distribution $q(\x)$. This method simplifies the learning process by focusing on modelling straighter probability paths, making the vector fields easier to learn and leading to faster generation time by reducing the number of steps required to generate high-quality samples. These probability paths are defined as an Ordinary Differential Equation (ODE) in Eq. (\ref{eq:ode}).

\begin{align}
\tfrac{d}{dt}\phi_t(\x)
& = \boldsymbol{v}_t(\phi_t(\x))
\text{;}
\qquad
\phi_0(\x)
= \x
\text{.}
\label{eq:ode}
\end{align}
where vector field $\boldsymbol{v}_t: [0, 1] \times \real^d \rightarrow \real^d$, generates the flow $\phi_t: [0, 1] \times \real^d \rightarrow \real^d$, which can be used to construct these probability density paths $p_t(\x)$. In practice, we learn the vector field $v_t$ with a neural network and minimise the expectation of MSE between the predicted $v_t$ and formulated vector field conditioned on samples from the data $\x_1$
\begin{align}
\mathcal{L}_{\mathrm{CFM}} (\theta)
& = \mathbb{E}_{t, q(\x_1),p_t(\x|\x_1)}\Vert \boldsymbol{u}_t(\x\vert \x_1) - \boldsymbol{v}_t(\x; \theta) \Vert^2.
\end{align}

We used Optimal Transport Conditional Flow Matching (OT-CFM) formulation from \cite{lipman2023flow} to construct the conditional vector field $\boldsymbol{u}_t$.


\subsection{Audio Language Models}

Earlier Audio LLMs involved integrating text-based LLMs with other audio foundation models \cite{huang2024audiogpt, liang2024wavcraft}, later evolving to combine text and audio within LLMs~\cite{deshmukh2023pengi, chu2023qwen, gong2023ltuas, kong2024audio} to enhance audio reasoning and comprehension. Previously, smaller language models had been employed for audio generation \cite{liu2024audioldm, agostinelli2023musiclm, kreuk2022audiogen, borsos2023audiolm}, though they lacked reasoning capabilities. Listen Think Understand (LTU) \cite{gong2023listen} introduced fine-tuning an LLM using continuous audio features for comprehension tasks, but lacked the capability for audio generation.\cite{nguyen2024spirit} introduced a speech tokenization approach for training LLMs for TTS and ASR by employing HuBERT and training the entire LLaMa-2 7b model. This approach, however, can be extremely computationally demanding and is hard to scale beyond the tasks it is trained on. \cite{hao2023boosting} attempted to enhance LLMs for TTS but concluded that LoRA was not well-suited for this task. Inspired by their findings, we sought to assess LoRA's effectiveness in general-purpose audio generation. Our proposed pipeline presents the first truly unified approach towards audio comprehension and generation.

\subsection{LoRA finetuning} 
LoRA (Low-Rank Adaptation) \cite{hu2021lora} provides an efficient method for fine-tuning large language models by introducing trainable low-rank matrices. In this approach, the original model weights $W_{orig}$ are kept frozen, and new learnable $\Delta W = A . B$ are added, where $A$ and $B$ are low-ranked. This results in the updated weight $W_{new} = W_{orig} + \Delta W$. LoRA is particularly advantageous for fine-tuning foundation models because it only trains a small fraction of the model's parameters, typically around 2-3\% of the total number of parameters reducing the computational cost significantly.

\begin{table}
\centering
\caption{Reconstruction errors $(\downarrow)$ for different datasets using VQ + FM and VQ + MSE approaches.}
\begin{tabular}{llcc}
\toprule
\textbf{Dataset} & \textbf{Type of audio} & \multicolumn{1}{c}{\textbf{VQ + FM}} & \multicolumn{1}{c}{\textbf{VQ + MSE}} \\
\midrule
ESC50           & General audio  & \textbf{0.6447} & 0.6532 \\
VCTK            & Speech         & 0.7435 & \textbf{0.7330} \\ 
FMA             & Music          & \textbf{0.5197} & 0.7150 \\ 
\midrule
\textbf{Average} &                & \textbf{0.6359} & 0.7004 \\ 
\bottomrule
\end{tabular}
\label{table:reconstruction_errors}
\vspace{-8pt}
\end{table}

\section{Method}
Training a causal LLM involves tokenizing the input text to maximize the conditional probability of each token at timestep $t$, given all previous tokens up to $t-1$. Formally, for a sequence of token $x_{1:T}$ we maximise $P \left(x_t \vert x_{<t} ; \theta \right)$ using cross entropy for all $1 < t \leq T $.
We downsampled the raw waveforms to reduce computational complexity and fixed the size of waveforms to 10 seconds and utilized a pretrained Autoencoder \cite{evans2024stable} to reduce the data size and make tokenization feasible. The waveform is passed through a convolution-based encoder which takes a 10-second stereo waveform at 44.1 kHz sampling rate and generates a representation $z \in \mathbb{R}^{215 \times 64}$ and reconstructs back using a decoder. We use this $z$ as our reconstruction target for our audio tokenizer module. 
For audio tokenization, we implement a causal tokenization strategy inspired by \cite{ge2023making}, which introduces a left-to-right bias in the representations, mirroring the sequential properties found in language. Instead of using a traditional decoder that minimizes the Mean Square Error (MSE) of reconstruction, as in regular VQ-VAE, we leverage conditional flow matching with Diffusion transformers (DiT) to reconstruct $\hat{z}$ as shown in Figure \ref{fig: quantizer architecture}. 

\begin{figure}[t]
    \centering
    \begin{subfigure}[b]{\columnwidth}
        \centering
        \includegraphics[width=\columnwidth]{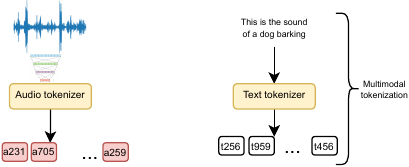}
        \caption{Audio and text tokenization}
        \label{fig:fig1}
    \end{subfigure}
    
    \vskip\baselineskip
    
    \begin{subfigure}[b]{\columnwidth}
        \centering
        \includegraphics[width=\columnwidth]{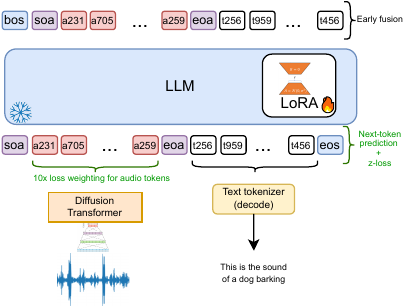}
        \caption{LLM training and generation process, where the losses highlighted in green are applied only during training, with additional tokens for the start of audio (soa) and end of audio (eoa).} 
        \label{fig:fig2}
    \end{subfigure}
    \caption{Overall pipeline for multimodal LLM}
    \label{fig: architecture overview}
    \vspace{-15pt}
\end{figure}

For the LLM, we utilize Vicuna 1.5 7b \cite{zheng2024judging} which is a fine-tuned variant of LLaMa 2 \cite{touvron2023llama} as the base model and incorporate LoRA adapters into all layers as shown in Figure \ref{fig: architecture overview}. Additionally, we resize the input and output embeddings to include audio tokens, training only these newly added embeddings while keeping the text embeddings frozen. Furthermore, we introduce beginning-of-audio and end-of-audio tokens into the vocabulary for handling audio tokens. During the pretraining stage, we utilized the audio captioning subset, using only the tokenized audio and caption text. We randomly swapped these as pairs of $\langle text, audio \rangle$ and $\langle audio, text \rangle$. During the fine-tuning stage, we utilised all the available datasets and used Vicuna's instruction template with added audio tokens. We applied 10x weighting on the new audio tokens and trained to minimise cross entropy with z-loss regularisation\cite{chowdhery2023palm}.

\section{Experiments}

\begin{figure*}
    \centering
    \includegraphics[]{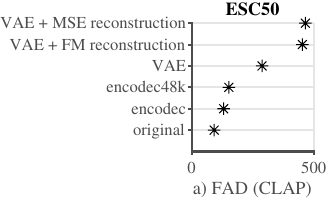}
    \includegraphics[]{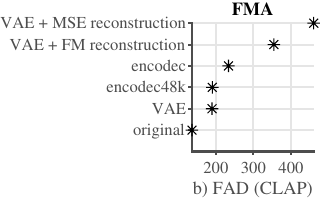}
    \includegraphics[]{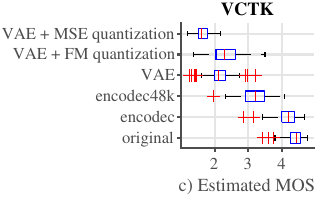}
    \caption{Audio quantization performance for held-out datasets in terms of FAD and estimated MOS.}
    \label{fig:audio quantization evaluation}
    \vspace{-8pt}
\end{figure*}

\subsection{Audio Tokenization}
For encoder, we used a feedforward transformer with causal attention, with 12 blocks, attention dimension of 64, and hidden dimension of 768. The codebook consisted of 8196 entries, each with 64 dimensions. The DiT decoder has a similar transformer block but without causal attention and added timestep embeddings of 256 dimensions. To assess the effectiveness of the Vector Quantization \textbf{(VQ)} with Flow Matching \textbf{(FM)}, we trained an identical architecture using Mean Square Error \textbf{(MSE)} loss while keeping the timestep input in DiT to constant zero. We trained both architectures for 75 epochs using AudioSet \cite{gemmeke2017audioset}, Clotho v2 \cite{lipping2019crowdsourcing}, VGG Sound \cite{chen2020vggsound}, FSD50k \cite{fonseca2021fsd50k}, BBCSoundEffects, SoundBible datasets on 4 Nvidia A6000s (48 GiB) with the AdamW optimizer set to a learning rate of 1e-4. For evaluation, we selected 100 samples each from the ESC50 \cite{piczak2015esc50}, VCTK \cite{veaux2016vctk}, and FMA \cite{defferrard2016fma} datasets to report the reconstruction error. In addition, for the VCTK dataset, we provided the predicted Mean Opinion Score for the reconstructed speech. Finally, to evaluate semantic relationship retention post-reconstruction, we used fadtk~\cite{gui2024adapting} with CLAP embeddings~\cite{elizalde2022clap} to obtain Fréchet Audio Distance (FAD) for the ESC50 dataset, using the remaining 1,900 samples as the reference, and for the FMA dataset, using the FMA-Pop subset as the reference, as outlined in \cite{gui2024adapting}. We also compared the FAD scores with \textbf{original} waveforms, waveforms reconstructed by Autoencoder of \cite{evans2024stable} \textbf{(VAE)}, EnCodec \cite{defossez2022high_encodec} both 24kHz \textbf{(encodec)} and 48kHz \textbf{(encodec-48k)}.

\subsection{Audio LLM}
For pretraining, we set LoRA rank to 64 and alpha to 128 and fine-tuned the model on $\approx$ 2M text-audio pairs with 900k unique audio samples from the closed caption subset of OpenAQA \cite{gong2023ltuas}, WavCaps, and FMA, totaling about 2500 hours of audio. We pretrained on 4 Nvidia A6000s (48 GiB) for 3 epochs. For full fine-tuning, we used the entire OpenAQA dataset containing $\approx$ 5M text-audio pairs and around 1.9M unique audio samples
and continued training on 24 Nvidia A100 (80 GiB)  for 30 epochs. We evaluated captioning  capabilities using SPICE \cite{anderson2016spice} and FENSE \cite{zhou2022can_fense} metrics from the  aac-metrics\footnote{\href{https://github.com/Labbeti/aac-metrics}{https://github.com/Labbeti/aac-metrics}} repository.

\section{Results and Discussions}

Table \ref{table:reconstruction_errors} compares the reconstruction errors of audio quantizers. Our VQ + FM method outperforms the VQ + MSE approach across general audio events and music, even though VQ + MSE was explicitly optimized to minimize MSE. Notably, the reconstruction error for the FMA dataset is significantly lower with the probabilistic model, highlighting the over-averaging issue that deterministic methods encounter when modelling VAE representations, especially those from \cite{evans2024stable}, which was primarily trained on music datasets. 
Figure \ref{fig:audio quantization evaluation} shows FAD scores for the ESC50 and FMA evaluation sets, where VAE + FM consistently captures the semantic relationships in audio more effectively than its MSE counterpart. It performs inferior to other codecs because of its ultra-low bitrate of \textbf{0.23 kbps} compared to VAE's 44 kbps, 24 kbps for encodec, and 6 kbps for encodec-48k. We also present the MOS scores after reconstructing samples from the VCTK dataset, concluding that speech is the most adversely affected, presumably due to the lack of speech data during the VAE training~\cite{evans2024stable}.

\begin{table}[h]
\centering
\caption{Audio captioning performance}
\begin{tabular}{lcccccc}
\toprule
\textbf{Model} & \multicolumn{2}{c}{\textbf{Clotho}} & \multicolumn{2}{c}{\textbf{AudioCaps}} & \multicolumn{2}{c}{\textbf{ESC50}} \\
& F $(\uparrow)$ & S $(\uparrow)$ & F $(\uparrow)$ & S $(\uparrow)$ & F $(\uparrow)$ & S $(\uparrow)$ \\
\midrule
 \textbf{LTU} & 0.41 & 0.08 & 
 0.35 & 0.04 & 0.44 & 0.14 \\
 \textbf{\architectureAbbrevationOnly} & 0.30 & 0.07 & 0.40 & 0.09 & 0.19 & 0.03 \\
\bottomrule
\end{tabular}
\label{table:comparison}
\end{table}

Table \ref{table:comparison} shows \architectureAbbrevationOnly\ performing competitively against LTU despite training on the same dataset at much lower bitrate than LTU's  76.8kpbs. 
However, SPICE (S) and FENSE (F) may penalize LTU's larger vocabulary, indicating the need for better captioning metrics. 
To investigate music comprehension, we tested our model on the GTZAN dataset \cite{gtzan2002musical}, where \architectureAbbrevationOnly\ correctly described 998 out of 999 files as music (e.g., "A harp is being played", "A drum loop is playing"). Table \ref{table: genre classification} summarizes results for instructing \architectureAbbrevationOnly\ to guess music genre. Accuracy was counted if the correct class was included in the model's predicted list. Note that LM-MSN consistently misclassified blues, country, disco, and reggae, mostly predicting them as folk or singer-songwriter.

\begin{table}[ht]
\centering
\caption{Genre Classification performance.}
\begin{tabular}{@{}ccc@{}}
\toprule
    Genre & Accuracy (\%) &    Most frequent output \\
\midrule
classical &    63.00 &               classical \\
  hip-hop &    76.00 &                 hip-hop \\
     jazz &    12.12 & folk, singer-songwriter \\
    metal &     0.00 &            rock, garage \\
      pop &    38.00 &                 hip-hop \\
     rock &    44.00 & folk, singer-songwriter \\
\bottomrule
\end{tabular}
\label{table: genre classification}
\end{table}

While we observe emerging generative abilities, \architectureAbbrevationOnly\ falls short compared to specialized audio synthesis models. Thus, we did not perform a formal evaluation but uploaded some samples on a webpage\footnote{\href{https://shivammehta25.github.io/LM-MSN}{https://shivammehta25.github.io/LM-MSN}}. This limitation may stem from the small dataset and bias in VAE representations, aligning with conclusions from \cite{hao2023boosting} for speech synthesis. Additionally, further research is needed to better understand neural scaling laws with LoRA and fine-tuning.

\section{Conclusion}

In this work, we introduced a pipeline for unifying audio comprehension and generation by utilizing pretrained text-only large language models (LLMs) and a novel audio tokenization method. By employing a Variational Quantization mechanism paired with Conditional Flow Matching, we were able to compress audio to an ultra-low bitrate, enabling the training of an audio language model within the LLM framework. Our evaluation demonstrated that the VQ + FM model outperformed deterministic approaches such as VQ + MSE. Further, we fine-tuned a pretrained text-based LLM using Low-Rank Adaptation (LoRA) for audio comprehension and generation tasks. Our findings suggest the potential of integrating audio and text modalities within LLMs using an early fusion approach, while also underscoring the importance of larger, more diverse datasets and improved evaluation metrics to fully unlock the capabilities of truly multimodal models.

\bibliographystyle{IEEEbib_abbrev}
\bibliography{IEEEabrv,references}

\end{document}